\documentclass[review,3p]{elsarticle}

\usepackage{hyperref}
\usepackage{longtable}
\usepackage{textcomp}
\usepackage{bm}
\usepackage{amsmath}
\usepackage{amsthm}
\usepackage{amssymb}
\usepackage{mathrsfs}
\usepackage{amsfonts}
\usepackage{color}
\usepackage{multirow}
\usepackage{cases}
\usepackage{graphicx}
\usepackage{tabularx}
\usepackage{mathrsfs}
\usepackage{mathdots}
\usepackage{algorithm}
\usepackage{algorithmic}
\usepackage{epstopdf}
\usepackage{setspace}
\newtheoremstyle{myplain}
  {\topsep}   
  {\topsep}   
  {\itshape\doublespacing}  
  {0pt}       
  {\bfseries} 
  {.}         
  {5pt plus 1pt minus 1pt} 
  {}
\theoremstyle{myplain}
\newtheorem{thm}{Theorem}[section]
\newtheorem{lem}{Lemma}[section]
\newtheorem*{prf}{Proof}
\newdefinition{rmk}{Remark}
\journal{Journal of \LaTeX\ Templates}

\begin{document}

\begin{frontmatter}

\title{Harmonic Retrieval with $L_1$-Tucker Tensor Decomposition\tnoteref{fund}}
\tnotetext[fund]{This work was supported by the National Natural Science Foundation of China (Grant No. 11771244, 11871297, 12171271).}


\author[mymainaddress]{Zhenting~Luan}
\author[mymainaddress]{Zhenyu~Ming}
\author[mysecondaryaddress]{Yuchi~Wu}
\author[mysecondaryaddress]{Wei~Han}
\author[mysecondaryaddress]{Xiang~Chen}
\author[mysecondaryaddress]{~Bo~Bai}
\author[mymainaddress]{Liping~Zhang\corref{mycorrespondingauthor}}
\cortext[mycorrespondingauthor]{Corresponding author}
\ead{lipingzhang@tsinghua.edu.cn}

\address[mymainaddress]{Department of Mathematical Sciences, Tsinghua University, Beijing 100084, China}
\address[mysecondaryaddress]{Theory Lab, Central Research Institute, 2012 Labs,  Huawei  Technologies Co., Ltd, Hong Kong.}

\begin{abstract}
Harmonic retrieval (HR) has a wide range of applications in the scenes where signals are modelled as a summation of sinusoids. Past works have developed a number of approaches to recover the original signals. Most of them rely on classical singular value decomposition, which are vulnerable to unexpected outliers. In this paper,  we present  new decomposition algorithms of third-order complex-valued tensors with $L_1$-principle component analysis ($L_1$-PCA)  of complex data and apply them to a novel random access HR model in  presence of outliers.  We also develop a novel subcarrier recovery method for the proposed model. Simulations are designed to compare our proposed method with some existing tensor-based  algorithms for HR. The results demonstrate the outlier-insensitivity of the proposed method.\end{abstract}


\end{frontmatter}


\section{Introduction}
Harmonic retrieval (HR)  is a vital problem in signal processing and has been applied to various areas, e.g.,  acoustic identification  \cite{advanceP2}, digital communication \cite{advanceP3.1} and biomedical processing  \cite{advanceP3.2}. The time-domain signals in HR model are naturally described as the summation of  sinusoids. The goal of the HR problem is to retrieve the original subsignals from received noisy signals.

Researchers have proposed diverse methods to solve this problem. Kumaresan et al. \cite{cem18} presented a technique based on linear prediction and singular value decomposition (SVD) to recover the pole-zero model in noises and then devised another nonlinear optimization method \cite{cem19} to solve the same model. Papadopoulos et al. \cite{cem24} came up with an estimation scheme based on high-order statistics.  Kung et al. \cite{cem21} presented a subspace approximation method based on SVD. Papy et al. \cite{yuan} proposed a tensor  Tucker-decomposition method, which is combined with the frequency recovery technique called ESPRIT \cite{esprit} to solve HR problem. The tensor Tucker-decomposition method is a generalization of matrix principle component analysis (PCA) \cite{view}. {\color{black}However, this technique cannot directly adapt to sporadic outliers due to the obvious outlier-sensitivity of  Frobenius norm in the objective function in conventional Tucker-decomposition methods.  To cope with outliers or non-Gaussian noises, some robust approaches  \cite{robustfit,robusthr,irhosvd} for HR are presented.}

In this paper, to enhance robustness against outliers, we employ  complex $L_1$-PCA \cite{L1PCA,L1TD} in  Tucker-decomposition  method for HR.  $L_1$-PCA  robustify conventional PCA scheme against outliers and shows well outlier-resistance property in  online outlier rejection \cite{outlierrejection} and streetscape Internet of Things \cite{iot}.

{\color{black}The  main  contributions  of  this  work  are as follows:  First, we develop two $L_1$-Tucker decomposition algorithms based on $L_1$-PCA to cope with the outliers in HR and   convergence analyses are performed. Second, we present a novel subcarrier sorting method (SCSM) for the proposed random access HR model. Simulation results show that SCSM yields more accurate results than conventional ESPRIT method, and the proposed $L_1$-Tucker decomposition algorithms outperform some existing tensor-based  algorithms in terms of outlier robustness.}

The paper is organized as follows. In Section \ref{pre},  we introduce the basic HR model and  Tucker decomposition.  In Section \ref{HRal}, we reformulate the HR problem (\ref{initialmodel}) with a random access model, apply  $L_1$-PCA into Tucker decomposition algorithms for solving the model and propose novel method  for subcarrier recovery. In Section \ref{simul},  some numerical experiments are presented for comparison between our proposed method and other tensor-based methods for HR. Finally,  conclusions are drawn.

\section{Preliminaries}\label{pre}
\subsection{Notations and Operators}

For an integer $n$,  denote $[n]=\{1,2,\ldots,n\}$. An $m$th-order tensor $\mathcal{A}\in \mathbb{C}^{I_1\times\cdots \times I_m }$ is an $m$-way array of complex numbers, which can be written as $\mathcal{A}=(\mathcal{A}_{i_1\cdots i_m}), i_j \in [I_j], j\in [m].$ {\color{black}Some tensor-related concepts in this paper are listed as follows:}
\begin{itemize}
    {\color{black}
\item \textbf{$k$-mode product $\times_k$:} product between an $m$th-order tensor $\mathcal{A}\in \mathbb{C}^{I_1\times\cdots \times I_m }$ and a matrix $M_k\in \mathbb{C}^{R_k\times I_k}$, which is denoted as $(\mathcal{A}\times_k M_k)_{i_1\ldots i_{k-1}ji_{k+1}\ldots i_{m}}=\sum_{i_k=1}^{I_k}a_{i_1\ldots i_{k-1}i_ki_{k+1}\ldots i_m}M_{ji_k}$
for $\forall i_l\in [I_l], \forall l\in [m]$ and $l\neq k, \forall j\in [R_k]$.
\item \textbf{$k$-mode fiber and $k$-mode unfolding:} $k$-mode fiber is $I_k$-length vector $\mathcal{A}_{i_1\ldots i_{k-1}: i_{k+1}\ldots i_m}$. $k$-mode unfolding recorded as $\mathcal{A}_{(k)}$, which aligns all $k$-mode fibers of $\mathcal{A}$ as the columns of an $I_k\times\prod_{l\neq k}I_l$ matrix with the lexicographical order of $i_1,\ldots,i_{k-1},i_{k+1},\ldots,i_m$. Then   the $k$-mode product can be reformulated as $(\mathcal{A}\times_k M_k)_{(k)}=M_k\mathcal{A}_{(k)}.$

\item \textbf{$n$-rank \cite{view}:} for each $n\in [m]$,  $n$-rank of tensor $\mathcal{A}$ is denoted as $r_n=rank(col(\mathcal{A}_{(n)}))$. Moreover, $\mathcal{A}$ is called an $(r_1,\cdots,r_m)$-rank tensor.
\item  \textbf{frontal-slice-Hankel (fs-Hankel) tensor:} we call a tensor $\mathcal{H}\in \mathbb{C}^{I_1\times\cdots \times I_m}$ \textit{fs-Hankel tensor} if for $i_m$-th ($i_m\in [I_m]$) frontal slice of tensor $\mathcal{H}$, there exists an $(I_1\!+\!\cdots\!+\!I_{m-1}\!-\!m\!+\!2)$-dimensional generating vector $\mathbf{h}^{(i_m)}$ such that $\mathcal{H}_{i_1\ldots i_m}=\mathbf{h}^{(i_m)}_{i_1+\ldots+i_{m-1}-m+2}.$
}
\end{itemize}

The complex sign operator of matrix $A\in \mathbb{C}^{m\times n}$ is
$sgn(A) = (A_{ij}/|A_{ij}|)_{m\times n}.$ Assume SVD of $A\in \mathbb{C}^{m\times n}$ is formulated as $A=UDV^H,$
where $D$ is diagonal and $U, V$ are column-unitary. Then the orthogonal projection operator $unt(A)$ is defined as $unt(A)=UV^H.$

\subsection{The HR Model and Tucker Decomposition of Tensors}

In  HR problem, the receiver observes a series of discrete time-domain signals,  which are formulated as\begin{equation}\label{initialmodel}
\textbf{x}_n=\sum_{k=1}^{K} \textbf{c}_ke^{j\omega_k t_n}+\textbf{q}_n, \quad n=0,1,\ldots,N-1,
\end{equation}
where $\textbf{x}_n$ is the $n$-th signal sample, $K$ is the number of subcarriers,  $\omega_k, \textbf{c}_k$ are  $k$-th pulsation  and corresponding complex amplitude $c_k$, respectively. $N$ is the number of samples, $\textbf{q}_n$ are  noises, and $t_n=n\Delta t,$  where $\Delta t$ is the sampling time interval. We need to recover the pulsations  $\omega_k$ and corresponding complex amplitude $\textbf{c}_k$ for all $k\in [K]$.  Let $z_k=e^{j\omega_k \Delta t}$ be the $k$-th pole of the signal, then (\ref{initialmodel}) is expressed as\begin{equation}\label{model}
\textbf{x}_n=\sum_{k=1}^{K} \textbf{c}_kz_k^n+\textbf{q}_n, \quad n=0,1,\ldots,N-1.
\end{equation}
Tucker decomposition is a powerful method to solve (\ref{model}) since the signals can be naturally stacked in an Hankel tensor of which each unfold matrix has a Vandermonde form \cite{yuan}.

Define $\mathbb{U}^{m\times n}\!=\!\{X\!\in\!\mathbb{C}^{m\times n}|X^HX\!=\!I_n\}.$ The Tucker decomposition of $m$th-order tensor $\mathcal{X}$ is written as\begin{equation}\label{tucker}
\mathcal{X}\approx \mathcal{C}\times_1 M_1\cdots \times_m M_m,
\end{equation}
where $\mathcal{X}\in \mathbb{C}^{I_1\times \ldots \times I_m}, \mathcal{C}\in \mathbb{C}^{R_1\times \ldots \times R_m},$ and $M_i\in \mathbb{U}^{I_i\times R_i}, \forall i\in[m].$ $\mathcal{C}$ is called the \textit{core tensor}, and $i$-th factor matrix $M_i$ can be considered as the principal components of $i$-mode unfolding of $\mathcal{X}$. If $R_i=I_i, \forall i\in [m],$  there exist $\mathcal{C}\in \mathbb{C}^{I_1\times\ldots \times I_m}$ and unitary matrices $M_i$ such that two sides of (\ref{tucker}) are equal \cite{yuan8}. For generic tensor, when $R_i<I_i, \forall i\in [m],$ it is usually infeasible for equality (\ref{tucker}) to hold, but the smaller core tensor $\mathcal{C}$ can be deemed as a compression or a low-rank approximation of $\mathcal{X}$. Hence, the generalized Tucker decomposition is to find the best $(R_1,\cdots,R_m)$-rank approximation of $\mathcal{X}$, which can be modeled as\begin{equation}\label{approxmodel}
\mathop{\arg\min}_{\substack{\mathcal{C}\in \mathbb{C}^{R_1\times\ldots \times R_m} \\  M_i\in \mathbb{U}^{I_i\times R_i}, i\in [m]}} \lVert\mathcal{X}-\mathcal{C}\times_1 M_1\cdots \times_m M_m\rVert_F
=  \mathop{\arg\max}_{\substack{ M_i\in \mathbb{U}^{I_i\times R_i}, i\in [m] \\ \mathcal{C}=\mathcal{X}\times_1 M_1^H\cdots \times_m M_m^H}} \lVert\mathcal{X}\times_1 M_1\cdots \times_m M_m\rVert_F.
\end{equation}

\section{Harmonic Retrieval with $L_1$-Tucker Decomposition of Complex Tensors}\label{HRal}

\subsection{System Model and Tensor-based Method with ESPRIT}\label{multimodel}

Here we consider a random access multi-symbol HR  model close to  real communication scenario, which has an exponential structure similar to (\ref{initialmodel}) but transmits signal on discrete frequencies.  Assume there are $K\!-\!1$ available subcarriers with poles $z_k=e^{j2\pi k/K}, k\in [K\!-\!1]$ and $K_a$ transmitters each randomly choose a different subcarrier  to transmit symbols, respectively. The receiver observes the composite signals by sampling $Q$ continuous symbols. The time-domain samples are formulated as\begin{equation}\label{simulationmulti}
\textbf{x}_n^{(q)}=\sum_{k=1}^{K-1} a_k\textbf{c}_k^{(q)} z_k^n+\textbf{q}_n^{(q)}, \quad 0\leq n<N, q\in [Q],
\end{equation}
where $\textbf{x}_n^{(q)}$ is the $n$-th composite sample of $q$-th symbol, $a_k=1 ($or $0)$ indicates that the $k$-th subcarrier is active (or inactive), $\textbf{c}_k^{(q)}$ is the  $q$-th transmitted symbol on $k$-th subcarrier  and is chosen from the given constellation, e.g., QPSK, $\textbf{q}_n^{(q)}$ is additive white Gaussian noise and the sampling interval is set as $\Delta t=1$. The receiver should recover the active subcarriers and corresponding transmitted symbols without knowing activity of transmitters and subcarriers. Like in \cite{yuan}, we assume that $K_a$ is known in advance.

The HR model (\ref{simulationmulti}) can be solved by the tensor-based method introduced in \cite{yuan}. We sketch the main idea as follows. The samples can be stacked in a third-order fs-Hankel tensor $\bm{\mathcal{H}}\in \mathbb{C}^{I_1\times I_2 \times Q}$ which satisfies $I_1+I_2-1=N,$  i.e.,
\begin{equation}\label{Hnew}
\bm{\mathcal{H}}_{i_1i_2q}=\textbf{x}_{i_1+i_2-2}^{(q)}=\sum_{k=1}^{K-1} a_k\textbf{c}_k^{(q)}(z_k^{i_1-1}z_k^{i_2-1})+\textbf{q}_{i_1+i_2-2}^{(q)}, \forall q\in [Q], i_j\in [I_j], j=1,2.
\end{equation}
Here $I_1$ and $I_2$ should be larger than $K_a$. Define the Vandermonde vector set as\begin{equation}\label{W}
W^{(j)}=\{w_i^{(j)}=\frac{1}{I_j}(1,z_i,\ldots,z_i^{I_j-1})^T | i\in [K-1]\}, \qquad W_{\text{active}}^{(j)}=\{w_i^{(j)} | a_i=1\}, \qquad j=1,2.
\end{equation}
In the noise-free case, the column subspace of $\bm{\mathcal{H}}_{(j)}$ is exactly spanned by $W_{\text{active}}^{(j)}$. The $(K_a,K_a,Q)$-rank Tucker decomposition of $\bm{\mathcal{H}}$,\begin{equation}\label{otherdecomp}
\bm{\mathcal{H}}\approx\widetilde{\mathcal{C}}\times_1 \widetilde{\textbf{U}}_1\times_2 \widetilde{\textbf{U}}_2 \times_3 \widetilde{\textbf{Q}},
\end{equation}
estimates the column subspace of $\bm{\mathcal{H}}_{(j)}$ with factor matrix $\widetilde{\textbf{U}}_j\in \mathbb{U}^{I_j\times K_a}, j=1,2$. In terms of the immanent shift-invariance property of $span(W_{\text{active}}^{(j)})$, we solve the matrix equation
\begin{equation}\label{Q3}
\widetilde{\textbf{U}}_j(2:I_i,:)={\widetilde{\textbf{U}}}_j(1:I_i-1,:)\widetilde{\textbf{Z}}.
\end{equation}
to obtain the solution $\widetilde{\textbf{Z}}^*$, then all eigenvalues $\hat{z}_k,k\in [K_a]$ of $\widetilde{\textbf{Z}}^*$ are regarded as the estimations of poles of $K_a$ active subcarriers. By substituting the hard-decision of $\hat{z}_k, k\in [K_a]$ into (\ref{simulationmulti}) and solving the linear system of $\textbf{c}_k^{(q)}$, we can obtain the estimations of symbols and the model (\ref{simulationmulti}) is completely recovered. The above subcarrier recovery scheme is called ESPRIT \cite{esprit}.

\subsection{$L_1$-Tucker Decomposition of Third-Order Complex Tensor}\label{l1al}

The most important step of the subspace method introduced in the last subsection is to compute the approximate Tucker decomposition (\ref{otherdecomp}).  Performance of a Tucker decomposition algorithms depends on its accuracy of separating initial signals (\ref{model}) from noises.  Higher-order  singular value decomposition (HOSVD) \cite{view226} and higher-order orthogonal iteration (HOOI)  \cite{view64} are two most popular algorithms for Tucker decomposition. Both two algorithms employ traditional PCA \cite{1933PCA} in the  iteration for each tensor mode to extract  principal components of the tensor-unfolding matrices, which is deemed as signal extraction from noises.

For any given matrix $\textbf{X}\in \mathbb{C}^{D\times N},$  classical PCA method solves the optimization problem,\begin{equation}\label{opt2}
\textbf{P}_{{Frob-opt}}=\mathop{\arg\max}_{\textbf{P}\in \mathbb{U}^{D\times K}} \lVert \textbf{P}^H\textbf{X} \rVert_F
\end{equation}
to obtain the first $K$ principal components of $\textbf{X}$, where $\lVert \textbf{A}\rVert_F=\sqrt{\sum_{i,j}|\textbf{A}_{i,j}|^2}$ is the Frobenius norm. {\color{black}However, due to the Frobenius-norm objective function, classical PCA is severely sensitive to heavy noises or outliers, which are inevitable in reality due to bursty interference in communication channels.  Thus we consider a more robust method, $L_1$-PCA  \cite{L1PCA}, to resolve this issue. }The $L_1$-PCA replaces the Frobenius-norm in  (\ref{opt2}) with the  $L_1$-norm operator $\lVert \textbf{A}\rVert_{1}=\sum_{i,j}|\textbf{A}_{ij}|$ and then gives as\begin{equation}\label{opt1}
\textbf{P}_{L_1-opt}=\mathop{\arg\max}_{\textbf{P}\in \mathbb{U}^{D\times K}} \lVert \textbf{P}^H\textbf{X} \rVert_{1}.
\end{equation}
Algorithm \ref{pca} \cite{L1PCA} reviews the convergent alternative iteration scheme for solving (\ref{opt1}).

\begin{algorithm}
 \caption{Algorithm for $L_1$-PCA$(\textbf{X},K,\textbf{P}^{(0)},\delta)$  \cite{L1PCA}}
 \begin{algorithmic}[1]\label{pca}
 \renewcommand{\algorithmicrequire}{\textbf{Input:}}
 \renewcommand{\algorithmicensure}{\textbf{Output:}}
 \REQUIRE $\textbf{X}\in \mathbb{C}^{D\times N}, K<$ rank$(\textbf{X}),\delta>0 ,k=1,$  initial matrix $\textbf{P}^{(0)}\in \mathbb{U}^{D\times K}$
 \ENSURE $\textbf{P}_{L_1-opt}$, the optimize solution of (\ref{opt1})
  \STATE $B^{(0)} = sgn(\textbf{X}^H\textbf{P}^{(0)}) \in \mathbb{C}^{N\times K}$
  \STATE $B^{(k)}\leftarrow sgn(\textbf{X}^Hunt(\textbf{X}B^{(k-1)}))$
 \WHILE{$|\rVert \textbf{X}B^{(k)}\rVert_*-\rVert \textbf{X}B^{(k-1)}\rVert_*|>\delta$ }
  \STATE $B^{(k+1)}\leftarrow sgn(\textbf{X}^Hunt(\textbf{X}B^{(k)}))$
  \STATE $k\leftarrow k+1$
  \ENDWHILE
 \RETURN $\textbf{P}_{L_1-opt}\leftarrow unt(\textbf{X}B^{(k)})$
 \end{algorithmic}
 \end{algorithm}

When the HR model (\ref{initialmodel}) is interfered by outliers, we can replace the subspace estimation step in the existing Tucker-decomposition methods, HOSVD and HOOI, with $L_1$-PCA to enhance robustness. To be consistent with the new PCA scheme, the objective function (\ref{approxmodel}) of approximate Tucker decomposition is adjusted in $L_1$-norm formula:
\begin{equation}\label{l1obj}
\mathop{max}_{\textbf{U}_i\in \mathbb{U}^{I_i\times K}, i=1,2,3} \lVert\bm{\mathcal{H}}\times_1\textbf{U}_1\times_2\textbf{U}_2\times_3\textbf{U}_3\rVert_{1}.
\end{equation}
Then we can naturally embed the Algorithm \ref{pca} into the subspace estimation step of HOSVD and obtain the Algorithm \ref{alTOTD} named as $L_1$-Third Order Tucker Decomposition ($L_1$-TOTD).  Algorithm \ref{alTOTD} also converges since it only utilizes $L_1$-PCA once for each mode of the given tensors.

 \begin{algorithm}
 \caption{$L_1$-TOTD}
 \begin{algorithmic}[1]\label{alTOTD}
 \renewcommand{\algorithmicrequire}{\textbf{Input:}}
 \renewcommand{\algorithmicensure}{\textbf{Output:}}
 \REQUIRE $\bm{\mathcal{H}}\in \mathbb{C}^{I_1\times I_2 \times I_3}, K>0 ,\delta>0, $ $\{\textbf{U}^{(0)}_i\in \mathbb{U}^{I_i\times K}\}$
 \ENSURE Factor matrices $\textbf{U}_i$ for each mode-$i$
 \FOR {$i=1,2,3$}
 \STATE $\textbf{U}_i\leftarrow L_1$-PCA$(\bm{\mathcal{H}}_{(i)},K,\textbf{U}^{(0)}_i,\delta)$
 \ENDFOR
 \RETURN $\{\textbf{U}_i\in \mathbb{C}^{I_i\times K}\}$
 \end{algorithmic}
 \end{algorithm}

Another classical Tucker decomposition algorithm,  HOOI,  also relies on PCA  during each sub-iteration for updating the factor matrices of each tensor mode. We also embed the $L_1$-PCA scheme into HOOI and gain the $L_1$-Third Order Orthogonal Iteration ($L_1$-TOOI) method as Algorithm \ref{alTOOI}. For instance, when the procedure runs to update the  $1$-mode factor matrix, it can be formulated as an $L_1$-PCA problem, \begin{equation}\label{HOOIPCA}
\textbf{U}_1\leftarrow\mathop{\arg\max}_{\textbf{U}\in \mathbb{U}^{I_1\times K}} \lVert \bm{\mathcal{H}}_{(1)}\times_1 \textbf{U}\times_2 \textbf{U}_2\times_3 \textbf{U}_3\rVert_1 =\mathop{\arg\max}_{\textbf{U}\in \mathbb{U}^{I_1\times K}} \lVert \textbf{U}^HH_1\rVert_1,
\end{equation}
where $H_1=\bm{\mathcal{H}}_{(1)}(\textbf{U}_3\otimes \textbf{U}_2)$ and $\otimes$ is the Kronecker product.
 \begin{algorithm}
 \caption{$L_1$-TOOI}
 \begin{algorithmic}[1]\label{alTOOI}
 \renewcommand{\algorithmicrequire}{\textbf{Input:}}
 \renewcommand{\algorithmicensure}{\textbf{Output:}}

 \REQUIRE $\bm{\mathcal{H}}\in \mathbb{C}^{I_1\times I_2 \times I_3}, K>0 ,\delta>0$
 \ENSURE Factor matrices $\textbf{U}_i$ for each mode-$i$
 \STATE Compute initial factor matrices $\{\textbf{U}_i^{(0)}\in \mathbb{U}^{I_i\times K}\}$ by Algorithm \ref{alTOTD}, $p=0$
 \WHILE{Not converging}
 \STATE $p\leftarrow p+1$
 \FOR {$i=1,2,3$}
 \STATE $H_i^{(p)}\leftarrow \bm{\mathcal{H}}_{(i)}(\otimes_{j=3,\ldots,i+1} \textbf{U}_j^{(p-1)} \otimes_{j=i-1,\ldots,1} \textbf{U}_j^{(p)})$
 \STATE $\textbf{U}_i^{(p)}\leftarrow L_1$-PCA$(H_i^{(p)},K,\textbf{U}_i^{(p-1)},\delta)$
 \ENDFOR
 \ENDWHILE
 \RETURN $\{\textbf{U}_i^{(p)}\in \mathbb{C}^{I_i\times K}\}$
 \end{algorithmic}
 \end{algorithm}

The convergence of  Algorithm \ref{alTOOI} is supported by the monotonicity of alternative iteration, which is illustrated in the following Theorem \ref{L1TD_conver}.

\begin{thm}\label{L1TD_conver}
For any $p\geq 1$, the value in (\ref{l1obj}) monotonically increases within the $p$-th iteration (Line 3-7) of Algorithm \ref{alTOOI} and between adjacent iterations, i.e.,
\begin{align}
 \lVert (\textbf{U}_1^{(p)})^H H_1^{(p)} \rVert_{1}&\leq \lVert (\textbf{U}_2^{(p)})^H H_2^{(p)} \rVert_{1}\leq  \lVert (\textbf{U}_3^{(p)})^H H_3^{(p)} \rVert_{1},\label{proofconv1} \\
\lVert (\textbf{U}_3^{(p)})^H H_3^{(p)} \rVert_{1} &\leq \lVert (\textbf{U}_1^{(p+1)})^H H_1^{(p+1)} \rVert_{1}. \label{proofconv2}
\end{align}

\end{thm}
\begin{prf}
The first inequality in (\ref{proofconv1}) is derived from  the following inequalities,
\begin{equation}\label{proof1inequal}
\lVert (\textbf{U}_1^{(p)})^H H_1^{(p)} \rVert_{1}   = \lVert (\textbf{U}_2^{(p-1)})^H H_2^{(p)} \rVert_{1}=\lVert (P_2^{(p,0)})^H H_2^{(p)} \rVert_{1}  \leq \lVert (P_2^{(p,k)})^H H_2^{(p)} \rVert_{1}=\lVert (\textbf{U}_2^{(p)})^H H_2^{(p)} \rVert_{1},
\end{equation}
where $P_2^{(p,0)}= \textbf{U}_2^{(p-1)}$ and $P_2^{(p,k)}=\textbf{U}_2^{(p)}$ are the input matrix and output result of $L_1$-PCA$(H_2^{(p)},K,\textbf{U}_2^{(p-1)})$, respectively. The  inequality sign in (\ref{proof1inequal}) is supported by the monotonically increasing of Algorithm \ref{pca} \cite{L1PCA}. The second inequality in (\ref{proofconv1}) can be verified in a similar manner. Therefore, the monotonic increasing property holds in each iteration. On the other hand, between two adjacent iterations, we have\begin{equation*}
\lVert (\textbf{U}_3^{(p)})^H H_3^{(p)} \rVert_{1}   = \lVert (\textbf{U}_1^{(p)})^H H_1^{(p+1)} \rVert_{1}=\lVert (P_1^{(p+1,0)})^H H_1^{(p+1)} \rVert_{1}  \leq \lVert (P_1^{(p+1,k)})^H H_1^{(p+1)} \rVert_{1}=\lVert (\textbf{U}_1^{(p+1)})^H H_1^{(p+1)} \rVert_{1},
\end{equation*}
where $P_1^{(p+1,0)}= \textbf{U}_1^{(p)}$ and $P_1^{(p+1,k)}=\textbf{U}_1^{(p+1)}$ are the input  matrix and output result of $L_1$-PCA$(H_1^{(p+1)},K,\textbf{U}_1^{(p)})$, respectively. Therefore, (\ref{proofconv2}) holds for any $p\geq 1$. \qed
\end{prf}

The monotonic increasing property, in conjunction with the fact that  (\ref{l1obj}) is upper bounded since its domain is compact, implies the convergence of Algorithm \ref{alTOOI}. However, as (\ref{l1obj}) is not convex, we cannot make sure that the Algorithm \ref{alTOOI}  converges to the global optimal solution for any initialization $\textbf{U}_i^{(0)},i=1,2,3.$ Like the initialization strategy for HOOI \cite{view,1998tucker}, we  firstly perform  $L_1$-TOTD and take the results as an initialization of $L_1$-TOOI, which helps the algorithm convergent to a better stationary point.

\subsection{Subcarrier Sorting  Method (SCSM) for Subcarrier Recovery}\label{polerecover}

{\color{black}In  presence of heavy noises and outliers, the factor matrices in the Tucker decomposition (\ref{otherdecomp})  have severe deviations, which  leads to apparent perturbations in solving (\ref{Q3}). We present a more robust subcarrier recovery method without solving (\ref{Q3}) to address this issue.}   Before stating the proposed method, we first prove the following lemma.
\begin{lem}\label{maxk}
Suppose $V=\{v_i\}_{i=1}^m\subset \mathbb{C}^n(n<m)$ is a set of unit-norm vectors satisfying that any $n$ vectors in V are linearly independent, and $\textbf{U}\in \mathbb{U}^{n\times k}.$ If there exists an index subset $I_0\subset [m]$ with $|I_0|=k$ such that the column space of $\textbf{U}$ is  spanned by $V_0=\{V_i, i\in I_0\}$, then \begin{equation}\label{poleseparate}
\lVert v_p^H\textbf{U}\rVert_F^2 < 1 = \lVert v_{i}^H\textbf{U}\rVert_F^2, \quad \forall p\notin I_0, \forall i\in I_0.
\end{equation}
\end{lem}
\begin{prf}
Let $\textbf{U}=(\textbf{U}_1,\ldots,\textbf{U}_k),$ then $\{\textbf{U}_i\}_{i=1}^k$ is a set of orthonormal bases of $span(V_0)$ and we expand it into complete orthonormal bases of $\mathbb{C}^n$, which is denoted as $\{\textbf{U}_i\}_{i=1}^n.$ Therefore, for any $i\in [m],$ we decompose $v_i$ into $v_i=\sum_{s=1}^n (\textbf{U}_s^Hv_i)\textbf{U}_s.$ Then $\sum_{s=1}^n \lVert \textbf{U}_s^Hv_i\rVert_F^2=\lVert v_i\rVert_F^2=1.$ Since $\forall i\in I_0,\forall s>k, \textbf{U}_s^Hv_i=0$, we have
$$\lVert v_{i}^H\textbf{U}\rVert_F^2=\sum_{s=1}^k \lVert \textbf{U}_s^Hv_i\rVert_F^2 =\sum_{s=1}^n \lVert \textbf{U}_s^Hv_i\rVert_F^2 =1.$$
For any $p\notin I_0,$  $v_p\notin span(V_0)$,  there exists $q>k$ such that $\textbf{U}_q^H v_p\neq 0$, and
\[
\pushQED{\qed}
\lVert v_{p}^H\textbf{U}\rVert_F^2=\sum_{s=1}^k \lVert \textbf{U}_s^Hv_p\rVert_F^2 < \sum_{s=1}^k \lVert \textbf{U}_s^Hv_p\rVert_F^2 + \lVert \textbf{U}_q^Hv_p\rVert_F^2\leq 1.\qedhere
\popQED
\]
\end{prf}

In the noise-free case, the Vandermonde set $W^{(1)}$ in (\ref{W})  and $\widetilde{\textbf{U}}_i=span(W_{\text{active}}^{(1)})\in \mathbb{C}^{I_1\times K_a}$ in (\ref{otherdecomp})  are in accordance with the conditions in Lemma \ref{maxk}. Hence,  we can  sort the elements of $S_{(1)}=\{\lVert w_i^{(1)H}\widetilde{\textbf{U}}_1\rVert_F\}_{i=1}^{K-1}$, and the largest $K_a$ elements    correspond exactly  to the $K_a$ active subcarriers. We call this method  subcarrier sorting method (SCSM). The performance comparison between SCSM and ESPRIT in Fig. \ref{polevsequation} illstrates robustness of SCSM in  presence of outliers.


\subsection{New Approach for Solving  (\ref{simulationmulti})}

Based on the algorithms proposed in Section \ref{l1al} and Section \ref{polerecover}, we obtain a novel approach to deal with   (\ref{simulationmulti})  as Algorithm \ref{harmonic}.

 \begin{algorithm}
 \caption{Algorithm for solving (\ref{simulationmulti})}
 \begin{algorithmic}[1]\label{harmonic}
 \renewcommand{\algorithmicrequire}{\textbf{Input:}}
 \renewcommand{\algorithmicensure}{\textbf{Output:}}

 \REQUIRE Signal samples $\{\textbf{x}_n^{(q)}\}_{n=0,q=1}^{N-1,Q}$ as (\ref{simulationmulti}), $I_1,I_2$ that satisfies $I_1+I_2=N+1,  K_a>0$
 \ENSURE  Active subcarriers $z_k$ and corresponding symbols $\textbf{c}_k^{(q)}$ for $k\in [K_a], q\in [Q]$

 \STATE Construct an fs-Hankel tensor $\bm{\mathcal{H}}\in \mathbb{C}^{I_1\times I_2\times Q}$ with the signal samples (
 \ref{simulationmulti})
 \STATE Obtain the factor matrices $\{\widetilde{\textbf{U}}_i\in \mathbb{C}^{I_i\times K}\}_{i=1,2}$ by $L_1$-TOTD or $L_1$-TOOI
 \STATE Run SCSM to search all $k_a$ active subcarriers  $\hat{z}_1,\ldots,\hat{z}_{K_a}$
 \STATE Substitute $\hat{z}_1,\ldots,\hat{z}_{K_a}$ into the initial model (\ref{simulationmulti}) and solve the linear system to obtain symbol estimations $\hat{\textbf{c}}_k^{(q)}, k\in [K_a],q\in [Q]$
 \end{algorithmic}
 \end{algorithm}
{\color{black}

The  first step in Algorithm \ref{harmonic} is to choose a proper scale $I_1\times I_2\times Q$ for  $\bm{\mathcal{H}}$. A necessary condition is that $I_1,I_2>K$ since (\ref{Q3}) requires that $\widetilde{\textbf{U}}_i^\uparrow$ is full-column-rank. Simulation results in \cite[Fig. 6]{yuan} advocated that slightly rectangular frontal slices may perform a little better than square slices.  Nevertheless, the best scale of such rectangular slices is empirical and  different in various scenes. For convenience, a common choice is to set $I_1=I_2$ as the scale of constructed fs-Hankel tensors,  as a compromise between convenience and optimality.}

\subsection{Cramer-Rao Lower Bound (CRLB)}\label{sectioncrlb}

Here we present the Cramer-Rao lower bound (CRLB) for estimations of $\textbf{c}_k^{(q)}$ and $z_k$. Assume the power of AWGN in (\ref{simulationmulti}) is $\sigma^2$. Let $\textbf{x}=\{\textbf{x}_n^{(q)}\}_{n=0,q=1}^{N-1,Q}$,  $\textbf{c}=\{\textbf{c}_k^{(q)}\}_{k=1,q=1}^K,Q$ and  $\textbf{z}=\{z_k\}_{k=1}^K$. The log-likelihood function is
\begin{equation}\label{loglike}
    \begin{aligned}
    \log p(\textbf{x}|\textbf{c},\textbf{z})&=\sum_{q=1}^Q\sum_{n=0}^{N-1}\log p(\textbf{x}_n^{(q)}|\textbf{c},\textbf{z}) \\
    &=\sum_{q=1}^Q\sum_{n=0}^{N-1}\left(-\log\pi\sigma^2-\sigma^{-2}|\textbf{x}_n^{(q)}-\sum_{k=1}^K\textbf{c}_k^{(q)}z_k^{n\Delta t}|^2\right)\\
    &=-NQ\log\pi\sigma^2-\sigma^{-2}\sum_{q=1}^Q\sum_{n=0}^{N-1}|\textbf{x}_n^{(q)}-\sum_{k=1}^K\textbf{c}_k^{(q)}z_k^{n\Delta t}|^2.
    \end{aligned}
\end{equation}
According to \cite[(13)]{crlb}, for any complex parameter $\theta$, we have
\begin{equation}\label{crlb}
\text{CRLB}(\theta)=\frac{-1}{E\left[\frac{\partial^2\log p(x,\theta)}{\partial\theta^*\partial\theta}\right]}.
\end{equation}
Then we have
\begin{equation}\label{ckcrlb}
    \text{CRLB}(\textbf{c}_k^{(q)})=\frac{-1}{E\left[\frac{\partial^2\log p(\textbf{x}|\textbf{c},\textbf{z})}{\partial \textbf{c}_k^{(q)*}\partial \textbf{c}_k^{(q)}}\right]}
    =\frac{1}{E\left[\sigma^{-2}\sum\limits_{n=0}^{N-1}(z_k^*)^{n\Delta t}z_k^{n\Delta t}\right]}
    =\sigma^{2}N^{-1}, k=1,2,\cdots,K,
\end{equation}
and
\begin{equation}\label{zkcrlb}
    \text{CRLB}(z_k)=\frac{-1}{E\left[\frac{\partial^2\log p(\textbf{x}|\textbf{c},\textbf{z})}{\partial z_k^*\partial z_k}\right]}
    =\frac{1}{E\left[\sigma^{-2}\sum\limits_{q=1}^Q\sum\limits_{n=0}^{N-1}(\Delta t)^2c_k^{(q)*}c_k^{(q)}\right]}
    =\frac{\sigma^2}{E(|c_k^{(q)}|^2)\Delta t^2NQ}.
\end{equation}

\section{Numerical Experiments}\label{simul}
{\color{black}
In this section, we compare the performance of Algorithm \ref{harmonic} with conventional tensor Tucker-decomposition  algorithms (HOOI, HOSVD) \cite{yuan} and robust tensor decomposition method called IR-HOSVD \cite{irhosvd}.

In the following simulations, there are $K\!-\!1\!=\!49$ available subcarriers. With randomly selected different subcarriers,  $K_a\!=\!6$ active transmitters simultaneously transmit $Q\!=\!12$ QPSK symbols in a frame, and the receiver samples $N\!=\!32$ time-domain samples for each symbol.  We assume that the samples in (\ref{simulationmulti}) are sparsely interfered by  outliers. The additive outliers are randomly and sparsely generated in signal samples and follow  the Gaussian distribution with large variance $\sigma_o^2\!=\!20.$

The performance of an algorithm is evaluated in terms of the root mean square error (RMSE) of subcarriers and symbols, which are defined as
\begin{equation}\label{rmse}
    \text{RMSE}_z=\sqrt{\mathbb{E}\left(|\hat{z}_k-z_k|^2\right)},~\text{RMSE}_c=\sqrt{\mathbb{E}\left(|\hat{\textbf{c}}_k^{(q)}-\textbf{c}_k^{(q)}|^2\right)},
\end{equation}
where $\hat{z}_k, \hat{\textbf{c}}_k^{(q)}$ are  estimations of $z_k$ and $\textbf{c}_k^{(q)}$, respectively.}

\begin{figure}[ht]
        \centerline{\includegraphics[width=14cm,height=5cm]{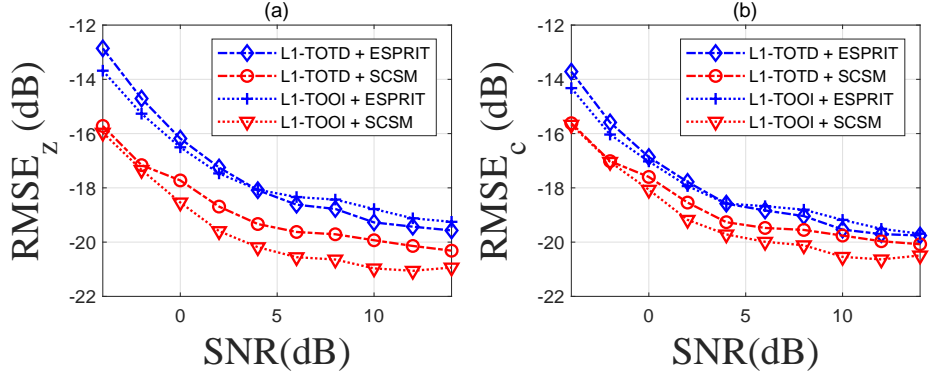}}
    \caption{Robustness of two subcarrier recovery methods: SCSM and ESPRIT. $10\%$ of signal samples are interfered by outliers.  (a) RMSE of subcarriers, (b) RMSE of symbols }\label{polevsequation}
\end{figure}

First we test the robustness of SCSM by comparing it with  ESPRIT, i.e., recovering subcarriers via solving equation (\ref{Q3}). The comparison results under a series of signal-to-noise ratio (SNR) are illustrated in
Figure \ref{polevsequation}.
The results shows that SCSM consistently yields better results. Specifically, SCSM shows better outlier-separation ability than ESPRIT under low SNR.

\begin{figure}[ht]
    \centerline{\includegraphics[width=14cm,height=5cm]{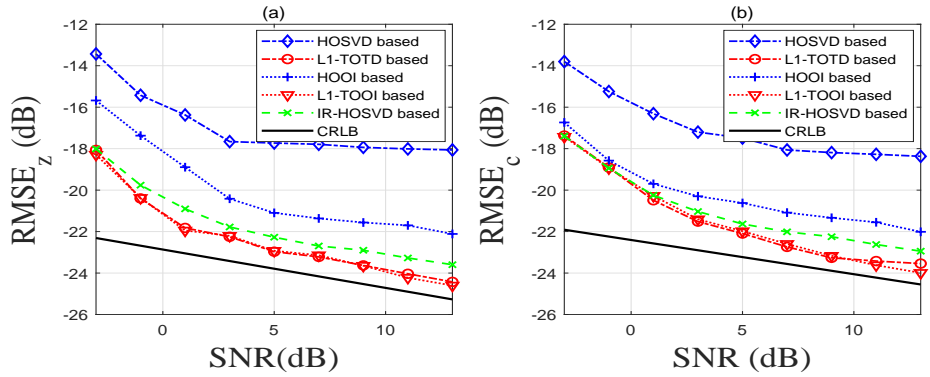}}
    \caption{Comparison between Algorithm \ref{harmonic} and HOSVD/HOOI/IR-HOSVD in  presence of outliers.  $5\%$ of signal samples are interfered by outliers.   (a) RMSE of subcarriers, (b) RMSE of symbols. }\label{allcomparison}
\end{figure}

Figure \ref{allcomparison}
reveals the comparison  among different tensor-based methods for solving (\ref{simulationmulti}) in terms of RMSE. The Cramer-Rao lower bound (CRLB) of (\ref{simulationmulti}) is given in \cite{l1hrarxiv}. Our proposed two algorithms significantly enhance the robustness of HOSVD and HOOI as  illustrated by the contrast of red and black curves in Fig. \ref{allcomparison}. Moreover, our method also outperforms IR-HOSVD, especially in high SNR.

\section{Conclusion}\label{conclusion}
Through leveraging  the weak outlier-sensitivity of $L_1$-norm, we introduce  $L_1$-Tucker decomposition algorithm to solve the harmonic retrieval  problem.   We also develop an SCSM scheme for subcarrier detection. Simulation results show that our algorithm exhibits  outlier-resistance property and outperforms  existing tensor-based approaches for HR problem.
\bibliography{lib}
\end{document}